\documentstyle[12pt]{article}

\begin{document}

\begin{titlepage}
\flushright{INRNE
\\TH-98/3}

\vskip 3.6truecm

\begin{center}
{\large \bf
SOME FUNCTIONAL SOLUTIONS OF \\[2mm] 
THE YANG-BAXTER EQUATION }
\end{center}

\vskip 1.0cm

\begin{center}
D.Ts. Stoyanov$^{\star}$ \\
\vskip 0.2cm
{\it Institute for Nuclear Research and Nuclear Energy \\
Boul.~Tsarigradsko chaussee 72, Sofia 1784, Bulgaria}
\vskip 1cm

\end{center}

\vskip 2.3cm

\rm
{\bf Abstract. }
A general functional definition of the infinite
dimensional quantum $R$-matrix satisfying the Yang-Baxter
equation is given. A procedure for extracting a finite dimensional
$R$-matrix from the general definition is demonstrated for the
particular cases of the group $O(2)$ and of the group of translations.
\vfill

\begin{flushleft}
\rule{5.1 in}{.007 in}\\

$^{\star}$ {\small e-mail address: dstoanov@inrne.acad.bg \\ }
\end{flushleft}

\end{titlepage}

\baselineskip=20pt plus 2pt minus 2pt 

\newpage

\section*{Introduction} 

The subject of this paper is the quantum $R$-matrix. This paper is a 
development of the idea proposed in the preprint \cite{c1}, 
and which is summarized in Section 1 below. This development includes a 
generalization and an announcement of some application of these matrices. 
This has a direct bearing on extensions of the $R$-matrix representations 
which we illustrate by constructing two simple examples using the 
simplest groups - $O(2)$ and the group of translations. 
We apply standard group theoretical methods. Note that 
quantum groups, which are the main sources of the usual $R$-matrix definition, 
are not considered here. In fact, we 
apply only the adjoint representation, and use the so-called canonical 
parameter, in which our expressions take a simpler form.

\section{The original model}
 
The subject of the present paper is the quantum
$R$-matrix with spectral parameter equal to zero 
(in the terminology of \cite{b1}). It is well known this $R$-matrix
satisfies the Yang--Baxter equation \cite{b2}: 
\begin{equation}
R_{12}R_{13}R_{23} = R_{23}R_{13}R_{12} \label{a1}
\end{equation}
Often one considers $R$ as nondegenerate finite
dimensional matrix acting on the tensor product
$V\otimes V$ of the finite-dimensional vector space $V$.
The notation $R_{ij}$
in eq.(\ref{a1}) signifies the matrix on
$V \otimes V \otimes V$
acting as $R$ on the $i$-th and $j$-th
components and as identity on
other component, e.g.,
$$ R_{12} = R \otimes {\rm id}$$

   The quantum $R$-matrix and the corresponding Yang--Baxter
equation are related to various problems of
theoretical and mathematical physics. The most significant
achievement is the observation that the solutions  of  the
Yang-Baxter equation are related to special algebras. These
are the deformed Lie algebras with comultiplication (Hopf
algebras), the so-called "quantum groups" \cite{b3}. Other new
algebras introduced in \cite{b4} are based in the exponential
solutions of the Yang--Baxter equation. These developments have
led to the  construction  of  some generalized expressions
for the $R$-matrix. However, one can hardly 
expect that these expressions exhaust all solutions of the
Yang--Baxter equation. The  new  multiparametric  nonstandard
solution of (\ref{a1}) presented in \cite{b5} is one such
example only.
     In the present paper we would like to reconsider the
problem of the exact solutions of Yang--Baxter equation  from
slightly
different point of view. We are going to present a new definition
for quantum $R$-matrix exploiting the idea of an operator acting
on a functional space. For this aim let us consider the space
$M$ of functions of two arguments defined on the direct product
$G\times G$, where
$G$ is an arbitrary Lie group. An arbitrary function from $M$ has
the form
\begin{equation}
f\equiv f(h;g) \label{a2} \end{equation}
where the variables $h$ and $g$
run over all elements of the group $G$ independently. Now  we 
define the following (Right and Left) operators: 
\begin{eqnarray}
R^{R}f(h;g) & = & w^{R}(h;g)f(h;hgh^{-1}) \label{a3} \\
R^{L}f(h;g) & = & w^{L}(h;g)f(g^{-1}hg;g) \label{a4}
\end{eqnarray}
Here the quantities $w^{R}(h;g)$
and $w^{L}(h;g)$
are the so-called "multiplicators" satisfying the equations
\begin{eqnarray}
w^{R}(h_{2};g)w^{R}(h_{1};h_{2}gh_{2}^{-1}) & = &
w^{R}(h_{2}h_{1};g)  \label{a5} \\
w^{L}(h;g_{2})w^{L}(g_{2}^{-1}hg_{2};g_{1}) & = &
w^{L}(h;g_{2}g_{1})  \label{a6}
\end{eqnarray}

   {\bf Theorem:} Let us denote with
$R^{A}_{ij}\,$,\ $A=R$ or $L$ and $i,j=1,2,3$, 
the operators acting on $i$-th and $j$-th argument of the function
$f(g_{1};g_{2};g_{3})$
defined on $G\times G\times G $  as $R^{A}$
from definitions (\ref{a3}) or (\ref{a4}) respectively.
Then the operators
$R^{A}_{ij}$ for every fixed $A$
satisfy the Yang--Baxter equation (\ref{a1}).

     The proof of this theorem is based on the properties
(\ref{a5}) and (\ref{a6})
of the multiplicators. Although their explicit form is not of
importance, we shall give a rather general
expression for these multiplicators. Let $T(g)$
be an arbitrary finite dimensional matrix representation
of the group $G$. Then it is easy to verify that the quantity
\begin{equation}
w^{R}(h;g)=\left(\frac{{\rm Sp}\, KT(hgh^{-1})}{{\rm Sp}\, KT(g)}\right)^d
 \label{a7}
\end{equation}
fulfills eq.(\ref{a5}), where with $K$
we have denoted an arbitrary but fixed matrix of the same
dimension, $d$ is real constant. We can write down similar
expression for the left multiplicator.

     Using the local coordinates in the Lie group G we can
give more effective form of our definitions, which is  useful
for the concrete applications. Here we shall choose the
coordinates usually called
"canonical parameters"  \cite{b6}.  In  these  parameters  the
representations of the local Lie  group $G$  have  strongly
exponential form (see the Appendix 1). Let us suppose that $G$ has $n$ 
parameters which we denote with
$\alpha_{\mu},\beta_{\nu},\dots\;\;\;(\mu,\nu=1,2,\dots,n)$.
Then the function $f(h;g)$
from (\ref{a2}) can be written in the coordinate form as
a function depending on $2n$ variables $f(\alpha_{\mu};\beta_{\nu})$,
where $\alpha_{\mu}$ and $\beta_{\nu}$
are parameters of the group elements $h$ and $g$
respectively.

In what follows we will consider the right
$R$-matrix
from definition (\ref{a3}) only. The corresponding
left matrices can be obtained analogously. If we denote with
$P^\nu_\mu(\alpha)$
the matrices of the adjoint representation of the group G, then the
parameters of the element $h^{-1}gh$
have the form
\begin{equation}
\beta'_\mu = P_\mu^\nu(\alpha_\rho)\beta_\nu \label{a8}
\end{equation}
This expression follows directly from the definition of the adjoint
representation. Moreover, the following identity
\begin{equation}
P_\mu^\nu(\alpha_\rho)\alpha_\nu = \alpha_\mu \label{a9}
\end{equation}
is fulfilled for the canonical parameters.
Then the definition (\ref{a3}) takes a new form
\begin{equation}
Rf(\alpha_\rho;\beta_\sigma) =
w(\alpha_\rho; \beta_\sigma)f(\alpha_\lambda;
P_\mu^\nu(\alpha_\tau)\beta_\nu)  \label{a10}
\end{equation}
Instead of the group elements in the arguments of the multiplicator we
substitute their parameters. Then eq.(\ref{a5}) can be rewritten
in the form
\begin{equation}
w(\alpha_{2\rho}; \beta_\sigma)w(\alpha_{1\rho};
P_\mu^\nu(\alpha_{2\omega})\beta_\nu)=
w\left( m_\rho(\alpha_{2\omega};\alpha_{1\tau}); \beta_\sigma \right)
\label{a11}
\end{equation}
where with $m_\rho(\alpha_{2\omega};\alpha_{1\tau})$
we denote the parameters of the element $h_2h_1$.
It is clear that the functions
$m_\rho(\alpha_{2\omega};\alpha_{1\tau})$
express the group multiplication law. In terms of the
group parameters we have chosen,
these  functions  satisfy the following simple equalities: 
\begin{equation}
m_\rho(P_\mu^\sigma(\alpha)\beta_\sigma;\alpha_\nu)=
m_\rho(\alpha_\nu; \beta_\mu)
\label{a12}
\end{equation}

       In general, $R$-matrix defined above is 
infinite-dimensional . As is well known for Lie
groups, the functions
$P^\nu_\mu(\alpha)\beta_\nu, w(\alpha_\rho;\beta_\sigma)$
and $m_\rho(\alpha_\mu;\beta_\nu)$
are smooth. According to the definition (\ref{a10}) of the operator $R$
the subspace $S$ of $M$ consisting of smooth functions is
invariant under the action of this operator. Using the basis of
orthonormal functions in the subspace $S$ we can obtain the matrix
form of the operator $R$.

Another invariant subspace one can obtain when $G$ is simple
compact group with structure constants ${C^{\mu\nu}}_\rho$. It is
well known that in this case
there exists a positive defined scalar product (Killing metric)
invariant with respect to the
adjoint representation of the group. In
particular if $\eta^{\mu\nu}$
is the Killing metric tensor, then the scalar square
\begin{equation}
\beta^2 = \eta^{\mu\nu}\beta_\mu\beta_\nu \label{a13}
\end{equation}
is invariant under the transformation (\ref{a8}). 

{\bf Remark:} The Killing metric tensor is defined as follows:
\begin{equation}
\eta^{\mu\nu} = \frac{-1}{n-1}
{C^{\mu\rho}}_\sigma {C^{\nu\sigma}}_\rho  \label{a17}
\end{equation}

\section{Generalization}

Now we want to generalize our considerations. First of all we start 
with the function from eq.(\ref{a2}) - $f(g;h)$. We modify our setting by 
replacing every variable by the tensor product of two variables. 
Thus, the new expression has the following form:
\begin{equation}
F(g_ig^1_k;h_nh^1_m)   \label{a18}
\end{equation}
where we write the group parameters 
$g_i\,$, $g^1_k\,$, $h_n\,$, $h^1_m\,$, 
($i,k,n,m$ $=$ $1,2,...,\xi=$ rank of $G$), 
instead of the group elements $g$, $g^1$, $h$, $h^1$, respectively, 
in order to have simpler calculations. 
Thus, the function $F$ is defined on the product\ $G\otimes G\times 
G\otimes G$. The tensor products arise since the factor 
$g_i\,$, $g^1_k\,$, respectively, will act on (or will be acted on by) 
the factor $h_n\,$, $h^1_m\,$, respectively. 

There are three different expressions that we take into account. 

We shall begin with the more difficult case that affects all 
variables which appear in the new function eq.(\ref{a18}). 
In particular, we shall need the following relation: 
\begin{equation}
F(g_i({{{(h^1)}^{-1}}g^1{h^1}})_k;({gh{g^{-1}}})_nh^1_m)=
F[({P^{-1}}(h^1))_k ^{k'}g_ig^1_{k'};P(g)_n^{n'}h_{n'}h^1_m] 
\label{a19}
\end{equation} 

Now we introduce the new $R$-matrix. To avoid complicated calculations 
we shall use the adjoint representation in the inner automorfism form.  
Then we define the new $R$-matrix as follows: 
\begin{equation}
RF(g_ig^1_k;h_nh^1_m)=
F[g_i({{(h^1)}^{-1}}g^1{h^1})_k;(gh{g^{-1}})_nh^1_m]
\label{a20}
\end{equation} 
so that we can use (\ref{a20}).  

Now we shall prove our proposal, i.e., we shall show that the operator $R$ 
in (\ref{a20}) satisfies the fundamental law for the
$R$-matrices (\ref{a1}).  
The left hand side of this formula takes the following form after action on $F$
$$R_{12}R_{13}R_{23}F(g_ig^1_k;h_nh^1_m;f_pf^1_q)= $$ 
$$=F[g_i({(f^1)}^{-1}{{(h^1)}^{-1}}f^1{(f^1)}^{-1}g^1f^1{(f^1)}^{-1}h^1g^1)_k;
(gh{g^{-1}})_n({(f^1)}^{-1}h^1f^1)_m;$$    
\begin{equation}
(ghf{h^{-1}}{g^{-1}})_pf^1_q]=    \label{a21}
\end{equation}
$$=F[g_i({(f^1)}^{-1}{(h^1)}^{-1}g^1h^1f^1)_k;(ghg^{-1})_n({(f^1)}^{-1}h^1f^1)_m;
(ghfh^{-1}g^{-1})_pf^1_q]$$     
This result is the same as the one that comes from the r.h.s. of the main
law (\ref{a1}). Actually the r.h.s. can be written
in the form (after action on $F$)
\begin{equation}   \label{a23}
R_{23}R_{13}R_{12}F(g_ig^1_k;h_nh^1_m;f_pf^1_q)
\end{equation}
The calculation which we have to do is analogous to the one we have done just above. 
After this the two expressions that we obtain from (\ref{a21}) and (\ref{a23}) 
are identical, i.e., eq.(\ref{a1}) is satisfied.  Using eq.(\ref{a19})
one can present everything in the adjoint representation. 

The two cases which remain, are analogous to those which we have considered 
the beginning of this paper (see eqs.(\ref{a3}) and (\ref{a4})), 
however, with tensor products as in the first case of this section. 
The $R$-matrices which give these formulae have the following form:
\begin{equation}   \label{a24}
R^RF(g_ig^1_k;h_nh^1_m)=F[g_ig^1_k;(ghg^{-1})_n(g^1h^1{(g^1)}^{-1})_m]
\end{equation}
\begin{equation}   \label{a25}
R^LF(g_ig^1_k;h_nh^1_m)=F[(h^{-1}gh)_i({(h^1)}^{-1}g^1h^1)_k;h_nh^1_m]
\end{equation}
The proof of the fact that that these operators satisfy eq.(\ref{a1})  
consists of applying several times the theorem of the previous section. 
In particular, we need to use eq.(\ref{a12}) twice, since we
have the tensor product setting with two variables

Finally, we can see that our statement is valid for many variables 
multiplied by tensor products. For example we write down
one of these expressions

$$R^RF(g^1_{i_1}g^2_{i_2}...g^n_{i_n};h^1_{k_1}h^2_{k_2}...h^n_{k_n})=$$
\begin{equation}
=F[g^1_{i_1}g^2_{i_2}...g^n_{i_n};
(g^1h^1{(g^1)}^{-1})_{k_1}(g^2h^2{(g^2)}^{-1})_{k_2}...
(g^nh^n{(g^n)}^{-1})_{k_n}]
\end{equation}
and so on for the remaining cases. Of course we add the multiplicator 
with appropriate arguments from eq.(\ref{a11}). However, we do not 
do this explicitly because it is a triviality.

\section{Example}

Now we will consider several examples. Let us begin with the simplest case. 
As first example we take the rotation group $O(2)$
\begin{equation}
|P_i^k(\alpha_3)|=\left(\matrix{\cos\alpha_3 &\sin\alpha_3 &0\cr
-\sin\alpha_3 &\cos\alpha_3 &0\cr 0 &0 &1\cr}\right)
\label{y1}
\end{equation}
as it rotates around the third axis. The function which we introduced in 
eq.(\ref{a10}) (without the multiplicator) has the following form:
\begin{equation}
Rf(\alpha_i;\beta_k)=f[\alpha_i;P_k^n(\alpha_3)\beta_n]  \label{ab1}
\end{equation}
(the indices take values $1,2,3$). In our case $\alpha_3$ and $\beta_3$ are not 
affected by the action of the group. 
That is why we consider the two dimensional case and $\alpha_3$ expresses
the rotation around the third axis. Then $\alpha_3$ is a parameter.
In our case (because the group is abelian) we can write omitting the 
function $f(\alpha_i;\beta_k)$ and using the external form of the bracket 
of this function. It follows
\begin{equation}
R_{i,k}^{m,n}=\delta_i^mP_k^n(\alpha_3)      \label{ab2}
\end{equation}
$$\delta_i^m=\cases {1 &if $i=m$\cr 0 &otherwise.\cr}$$
Now all indices up to the end of the example take values $1,2$.
To check that the last expression 
is a solution of the $R$-matrix 
rule we substitute (\ref{ab2}) in (\ref{a1}) to obtain: 
$$\delta_{i_1}^{j_1}P_{i_2}^{j_2}(\alpha_3)\delta_{i_3}^{j_3}
\delta_{j_1}^{k_1}\delta_{j_2}^{k_2}P_{j_3}^{k_3}(\alpha_3)
\delta_{k_1}^{n_1}\delta_{k_2}^{n_2}P_{k_3}^{n_3}(\beta_3)=$$
$$=\delta_{i_1}^{j_1}\delta_{i_2}^{j_2}P_{i_3}^{j_3}(\beta_3)
\delta_{j_1}^{k_1}\delta_{j_2}^{k_2}P_{j_3}^{k_3}(\alpha_3)
\delta_{k_1}^{n_1}P_{k_2}^{n_2}(\alpha_3)\delta_{k_3}^{n_3}$$
We see that after performing the summations we obtain the identity:
\begin{equation}
\delta_{i_1}^{n_1}P_{i_2}^{n_2}(\alpha_3)P_{i_3}^{n_3}(\alpha_3+\beta_3)=
\delta_{i_1}^{n_1}P_{i_2}^{n_2}(\alpha_3)P_{i_3}^{n_3}(\alpha_3+\beta_3)\ .  
\label{ab3}
\end{equation}
In this case 
dividing the interval $[0,2\pi]$ in $N$ parts (where $N$ is the number of external 
lines of regular polygon) we obtain the finite case:  
$${2\pi\over N} n\geq\alpha_3\geq {2\pi\over N} (n-1) \ .$$

Let us consider the next example. 
We shall use the case which we called "difficult". 
Here we shall use the noncompact - "translation" - group: 
$$\left(\matrix{
1 &0\cr
g_3 &1\cr}
\right)$$
This translation we obtain using the $SL(2,R)$ group (to which the above matrix 
belongs). The generators of the $SL(2,R)$ group take the form
\begin{equation}
q_1=\left(\matrix{1 &0\cr 0 &{-1}\cr}\right);\quad 
q_2=\left(\matrix{0 &{-1}\cr 0 &0\cr}\right);\quad 
q_3=\left(\matrix{0 &0\cr 1 &0\cr}\right)
\label{ab4}
\end{equation}
and we apply formula $(A.2)$ from Appendix 1 to obtain the following formula 
\begin{equation}
U_{g_3} q_k U_{-g_3}=q_jP^j_k(g_3)  \label{ab5}
\end{equation}
for the translation subgroup. 
The matrix in the r.h.s. takes the following form:
\begin{equation}
|P^j_k(g_3)|=\left(\matrix{1 &0 &2g_3\cr  
g_3 &1 &g^2_3\cr         \label{ab6}
0 &0 &1\cr}\right)
\end{equation} 
This matrix is not completely reducible. 
It has an invariant subspace: 
\begin{equation}
{\left(\matrix{1 &0 &2g_3\cr
g_3 &1 &g^2_3\cr           \label{ab7}
0 &0 &1\cr}\right)}{\left(\matrix{ h_1\cr h_2\cr 0\cr}\right)}=
\left(\matrix{h_1\cr {g_3h_1+h_2}\cr 0\cr}\right)
\end{equation} 
the complement to which is not invariant. 
That is why in our case there are two possibilities: with two dimensions and 
with three dimensions. We shall consider only the two dimensional case whereas
a reader can make the same in the three dimesional case.

The tensor product in the formula
\begin{equation}
f=f(g_ig^1_k;h_nh^1_m) \ ,\qquad i,k,n,m=1,2    \label{ab8}
\end{equation}
can be expressed in another way. Namely, 
the representation of the tensor product can be illustrated with the 
$4\times4$ matrices multiplied in the appropriate way. 
The first matrix has the following form: 
$$|P^{(-1)\beta}_\alpha(-h^1_3)|=
{\left(\matrix{        
1 &0\cr            
0 &1\cr}          
\right)}{\otimes}{\left(\matrix{1 &-h^1_3\cr            
                0 &1\cr}           
                \right)}=\left(\matrix{1 &0 &-h^1_3 &0\cr
0 &1 &0 &-h^1_3\cr
0 &0 &1 &0\cr
0 &0 &0 &1\cr}
\right)$$
and the second matrix is:                
$$|P_\alpha^\beta(g_3)|=
{\left(\matrix{
1 &g_3\cr
0 &1\cr}
\right)}{\otimes}{\left(\matrix{1 &0\cr
                0 &1\cr}
                \right)}=\left(\matrix{1 &g_3 &0 &0\cr
0 &1 &0 &0\cr
0 &0 &1 &g_3\cr 
0 &0 &0 &1\cr}
\right)$$ 
(All Greek indices take values from 1 up to 4.) 

Then we turn towards the variables that make the tensor product.               
We can represent the pair of two-dimensional variables 
also as a column of four arguments:
$$
U={\left(\matrix{
g_1g^1_1\cr
g_2g^1_1\cr
g_1g^1_2\cr
g_2g^1_2\cr}
\right)}=\left(\matrix{x_1\cr x_2\cr x_3\cr x_4\cr}\right)
$$
We do the same with the remaining variables.

{\bf Remark.}
Something similar we can do when we have the tensor product setting 
with more than two variables. 

Now we shall replace this with the new argument. The function which we introduced 
in eq.(\ref{ab8}) now takes the following form
\begin{equation}
f(g_ig^1_k;h_nh^1_m)=F(x_\theta;y_\eta) \label{ab9}
\end{equation}

Now we deal with the $R$-matrix. As in the rotation around the third axis,  
here the $R$-matrix is finite too. It has is the following form: 
\begin{equation}      
R_{\alpha\beta}^{\gamma\delta}={P^{(-1)\gamma}_\alpha(-h^1_3)} P_\beta^\delta(g_3)  
 \label{ab10}
\end{equation} 
It is more convenient to use for this $R$-matrix another equivalent form, 
which is the following:
\begin{equation}
R^{\gamma\delta}_{\alpha\beta}=(\delta^\gamma_\alpha-\delta^1_\alpha\delta^\gamma
_3h^1_3-\delta^2_\alpha\delta^\gamma_4h^1_3)
(\delta^\delta_\beta+\delta^1_\beta\delta^\delta_2g_3+
\delta^3_\beta\delta^\delta_4g_3)
\label{ab11}
\end{equation}
As expected this form aslo satisfies identically eq.(\ref{a1}). 
The calculation is in Appendix 2. 

 In the three dimensional case the calculation is a more painful task because it
is much longer. However, as a help for our reader we note that the product of 
two matrices of the kind given in (\ref{ab6})
\begin{equation}
{\left(\matrix{1 &0 &2g\cr g &1 &g^2\cr 0 &0 &1\cr}\right)}{\left(\matrix{
1 &0 &2h\cr h &1 &h^2\cr 0 &0 &1\cr}\right)}=
\left(\matrix{1 &0 &{2(g+h)}\cr {g+h} &1 &{(g+h)^2}\cr 0 &0 &1\cr}\right)
\label{ab12}
\end{equation}
clearly shows that (\ref{ab6}) indeed represents 
the one-dimensional translations.                      
\vskip 20pt
 {\bf Acknowledgment:} The author is grateful to Dr. V.K. Dobrev
for useful discussions.

\vskip 20pt
{\bf Appendix 1}

     Let $G$ be a $n$-parametric Lie group and
$\alpha_\mu, \;\;\;(\mu=1,2,\dots,n)$
be the canonical parameters of element $g\in G$.
According to \cite{b6} the canonical parameters are defined
with the help of the group structure functions
$S_\mu^\nu(\alpha_\rho)$
in the following way
$$S_\mu^\nu(\alpha_\rho)\alpha_\nu = \alpha_\mu \eqno(A.1)$$

     Let $T(g)$
be an arbitrary representation of the group $G$ with generators
$I^\mu$.
Then the adjoint representation matrices are defined as follows
$$T(\alpha)I^\mu T^{-1}(\alpha) = I^\nu P_\nu^\mu(\alpha)
\eqno(A.2) $$
     The main consequence of eq.(A.1) is that the operator
$T(g)$ can be written in the strougly exponential form:
$$T(\alpha) =\exp{\{i I^\mu\alpha_\mu\} }\eqno(A.3) $$

Combining formulas (A.2) and (A.3) we obtain that
$$\exp{\{i I^\mu\alpha_\mu\} }
\exp{\{i I^\nu\beta_\nu\} } =
\exp{\{i I^\nu P_\nu^\mu(\alpha)\beta_\mu\} }
\exp{\{i I^\rho \alpha_\rho\} } \eqno(A.4) $$
from which eq.(\ref{a12}) follows immediately.
$$m_\rho(P_\mu^\sigma(\alpha)\beta_\sigma;\alpha_\nu)=
m_\rho(\alpha_\nu; \beta_\mu ) \eqno(A.5) $$ 
The
case when $\alpha_\mu \equiv \beta_\mu$
the eq.(A.5) leads to the identity (\ref{a9}).

Finally the identity 
$$P_\mu^\rho(\beta)\delta_{\rho\nu}=P_\mu^{\mu\prime}(\alpha)
P_\nu^{\nu\prime}(\alpha)\delta_{\nu\prime\rho\prime}
P_{\mu\prime}^{\rho\prime}(P_\lambda^\sigma(\alpha)\beta_\sigma)
\eqno(A.6) $$ 
also follows from eq.(A.4)
when the latter is written for the adjoint representation.
This expression means that the matrix $P_\mu^\nu(\alpha)$
is an invariant tensor of second rank,
i.e. when
the generators $I^\mu$
are $(I^\mu)_\rho^\sigma = {C^{\mu\sigma}}_\rho \;\;
({C^{\mu\nu}}_\rho$
are the structure constant of group $G$ as above).

\vskip 20 pt
{\bf Appendix 2}

Here we prove that the $R$-matrix given in eq.(\ref{ab11})  
satisfies eq.(\ref{a1}). Substituting this $R$-matrix in eq.(\ref{a1}) we obtain:
$$(\delta_\alpha^\gamma-\delta_\alpha^1\delta_3^\gamma h^1_3- 
\delta_\alpha^2\delta_4^\gamma h^1_3)(\delta_\beta^\delta+ \delta_\beta^1
\delta_2^\delta g_3+\delta_\beta^3\delta_4^\delta g_3)\delta_\theta^\eta\cdot$$
$$\cdot(\delta_\gamma^{\gamma_1}-\delta_\gamma^1\delta_3^{\gamma_1} f^1_3-
\delta_\gamma^2\delta_4^{\gamma_1} f^1_3)\delta_\delta^{\delta_1}(\delta_\eta^{\eta_1}+
\delta_\eta^1\delta_2^{\eta_1} g_3+\delta_\eta^3\delta_4^{\eta_1} g_3)\cdot$$
$$\cdot\delta_{\gamma_1}^{\alpha_1}(\delta_{\delta_1}^{\beta_1}-\delta_{\delta_1}^1
\delta_3^{\beta_1} f^1_3-\delta_{\delta_1}^2\delta_4^{\beta_1} f^1_3)
(\delta_{\eta_1}^{\theta_1}+\delta_{\eta_1}^1\delta_2^{\theta_1} h_3+
\delta_{\eta_1}^3\delta_4^{\theta_1} h_3)=$$
$$=\delta_\alpha^\gamma(\delta_\beta^\delta-\delta_\beta^1\delta_3^\delta f^1_3-
\delta_\beta^2\delta_4^\delta f^1_3)(\delta_\theta^\eta+
\delta_\theta^1\delta_2^\eta h_3+\delta_\theta^3\delta_4^\eta h_3)\cdot
\eqno(B.1)$$
$$\cdot(\delta_\gamma^{\gamma_1}-\delta_\gamma^1\delta_3^{\gamma_1} f^1_3-
\delta_\gamma^2\delta_4^{\gamma_1} f^1_3)\delta_\delta^{\delta_1}
(\delta_\eta^{\eta_1}+\delta_\eta^1\delta_2^{\eta_1} g_3+
\delta_\eta^3\delta_4^{\eta_1} g_3)\cdot$$
$$\cdot(\delta_{\gamma_1}^{\alpha_1}-\delta_{\gamma_1}^1\delta_3^{\alpha_1} h^1_3-
\delta_{\gamma_1}^2\delta_4^{\alpha_1} h^1_3)(\delta_{\delta_1}^{\beta_1}+
\delta_{\delta_1}^1\delta_2^{\beta_1} g_3+\delta_{\delta_1}^3\delta_4^{\beta_1}
 g_3)\delta_{\eta_1}^{\theta_1}$$

After a simple calculation we have the following:
$$[\delta_\alpha^{\alpha_1}-\delta_\alpha^1\delta_3^{\alpha_1}(h^1_3+f^1_3)-
\delta_\alpha^2\delta_4^{\alpha_1}(h^1_3+f^1_3)]A_\beta^{\beta_1}(g_3;f^1_3)
\cdot$$
$$\cdot[\delta_\theta^{\theta_1}+\delta_\theta^1\delta_2^{\theta_1}(g_3+h_3)-
\delta_\theta^3\delta_4^{\theta_1}(g_3+h_3)]= \eqno(B.2)$$
$$=[\delta_\alpha^{\alpha_1}-\delta_\alpha^1\delta_3^{\alpha_1}(f^1_3+h^1_3)-
\delta_\alpha^2\delta_4^{\alpha_1}(f^1_3+h^1_3)]B_\beta^{\beta_1}(f^1_3;g_3)
\cdot$$
$$\cdot[\delta_\theta^{\theta_1}+\delta_\theta^1\delta_2^{\theta_1}(h_3+g_3)-
\delta_\theta^3\delta_4^{\theta_1}(h_3+g_3)]$$

It remains to show that the matrices $A_\beta^{\beta_1}(g_3;f^1_3)$ and 
$B_\beta^{\beta_1}(f^1_3;g_3)$ coincide. 
This is easy to prove and to show that: 
$$|A_\beta^{\beta_1}(f^1_3;g_3)|= 
|B_\beta^{\beta_1}(f^1_3;g_3)| = 
\left(\matrix{1 &{g_3} &{-f^1_3} &{-f^1_3g_3}\cr
0 &1 &0 &{-f^1_3}\cr 0 &0 &1 &{g_3}\cr 0 &0 &0 &1\cr}\right) \ . \eqno(B.3)$$

\end{document}